\documentclass[aps,reprint,article]{revtex4-1}
\usepackage{amsmath}
\usepackage{dcolumn}
\usepackage{graphicx}
\usepackage{float}

\begin{document}

\title{BCS-like critical fluctuations with limited overlap of Cooper pairs in FeSe}

\author{Huan Yang$^*$, Guanyu Chen, Xiyu Zhu, Jie Xing, and Hai-Hu Wen$^\dag$}

\affiliation{National Laboratory of Solid State Microstructures and Department of Physics, Collaborative Innovation Center of Advanced Microstructures, Nanjing University, Nanjing 210093, China}

\begin{abstract}
In conventional superconductors, very narrow superconducting-fluctuation regions are observed above $T_c$, because strong overlap of Cooper pairs occurs in a coherence volume $4\pi\xi^3/3$ with $\xi$ being the coherence length. In the bulk form of iron-chalcogenide superconductor FeSe, it is argued that the system may be located in the crossover region from Bardeen-Cooper-Schrieffer to Bose-Einstein condensation (BEC), where strong superconducting fluctuations are expected. In this respect, we carried out magnetization, specific heat and Nernst effect measurements on FeSe single crystals in order to investigate the superconducting fluctuation effect near $T_c$. The temperature range of diamagnetization induced by superconducting fluctuations seems very narrow above $T_c$. The temperature-dependent magnetization curves measured at different magnetic fields do not cross at a single point. This is in sharp contrast to the situation in many cuprate superconductors, where such a crossing point has been taken as a clear signature of strong critical fluctuations. The magnetization data can be scaled according to the Ginzburg-Landau fluctuation theory for a quasi-two-dimensional system. However the scaling result cannot be described by the theoretical function of the fluctuation theory due to the limited fluctuation regions. The specific heat jump near $T_c$ is rather sharp without the trace of strong superconducting fluctuations. This is also supported by the Nernst effect measurements which indicate a very narrow region for vortex motion above $T_c$. Associated with very small value of Ginzburg number and further analyses, we conclude that the superconducting fluctuations are vanishingly weak above $T_c$ in this material. Our results are strongly against the picture of significant phase fluctuations in FeSe single crystals, although the system has a very limited overlap of Cooper pairs in the coherence volume. This dichotomy provides new insights into the superconducting mechanism when the system is with a dilute superfluid density.
\end{abstract}

\maketitle

\section{INTRODUCTION}

In iron-pnictide/chalcogenide superconductors, the well-accepted pairing symmetry is the spin-fluctuation mediated nodeless $s\pm$ model, i.e., the superconducting gap changes sign between the hole and electron pockets \cite{Mazin,Kuroki}. In most systems, both structural and magnetic phase transitions appear in the temperature-doping ($T$-$x$) phase diagram. In addition, the antiferromagnetic (AFM) phase with an orthorhombic structure and a nematic electronic state can even coexist with the superconducting phase in the underdoped region in many systems \cite{StewartReview,WenReview}. Among iron-based superconductors, tetragonal FeSe has the simplest structure and its superconducting transition temperature $T_c$ is about 9 K \cite{WuMKPNAS}. Surprisingly, only the structural transition from tetragonal to orthorhombic was observed at $T_s \approx 90$ K in bulk FeSe without any trace of antiferromagnetic transition. Below $T_s$ a significant electronic anisotropy is induced \cite{SongCLScience}. Fermi surfaces revealed by angle resolved photoemission and quantum oscillations \cite{ARPES1,ARPES2,ARPES3,ARPES4,QuanOs1,QuanOs2} indicate the presence of hole and electron pockets with probably a strong non-degeneracy of the $d_{xz}$ and $d_{yz}$ orbitals. This leads to the breaking of the fourfold symmetry in the orthorhombic phase below $T_s$. Regarding the very shallow band top or bottom, it was argued that the Fermi energy is quite small and the effective charge carrier density in the material is very dilute. On the other hand, the superconducting transition temperature of FeSe can be easily enhanced to about 38 K at a pressure of 6 GPa, and there is also a pressure-induced magnetic transition dome in a wide pressure range \cite{ChengJGPressure,CanfieldPressure}. Stripe-type spin fluctuations, which are almost independent of pressure, have also been observed below $T_s$, reconciling FeSe with other iron-based superconductors \cite{YuWQPRL}. Furthermore, another interesting issue for superconductivity in FeSe is about the exact structure of superconducting gap, which is still under debate. The V-shaped tunneling spectra was observed near zero-bias by earlier scanning tunneling spectroscopy (STS) measurements on FeSe film \cite{SongCLScience}, which was argued as the evidence of a $d$-wave superconducting gap. This was corroborated by STS measurements on FeSe single crystals, and the tiny fully gapped states near the Fermi level was ascribed to the effect of the twin boundaries \cite{HanaguriPRX}. However, the nodeless gap feature is supported by the thermal conductivity \cite{TailleferPRL} and specific heat measurements \cite{FeSeCap,ChenGY2017}. Recently, a pair of nodeless sign-reversal gaps with extremely high anisotropy has been detected with very detailed Bogoliubov quasiparticle interference analysis based on the STS measurements \cite{DavisQPI}.

In a conventional superconductor, the superconducting pairing and condensation occur simultaneously at $T_c$, and the superconducting-fluctuation (SCF) region is very limited above $T_c$. This is because a great number of Cooper pairs lie within the coherence volume $4\pi\xi^3/3$, where $\xi$ represents the coherence length. In other words, Cooper pairs strongly overlap with each other in space. Actually the Cooper pairs in the coherence volume $4\pi\xi^3/3$ in conventional superconductors can be understood as highly entangled states of thousands of Cooper pairs. In high-$T_c$ cuprate superconductors, however, SCFs have been shown to be quite strong. It is argued that large Nernst signal or diamagnetic magnetization far above $T_c$ is a proof of the existence of strong phase fluctuations of preformed Cooper pairs above $T_c$ \cite{CuNernstReview}. However in another different view, some electronic order may compete with superconductivity above irreversible field or temperature \cite{CuprateReview}, which leads to the Nernst signal \cite{YBCONernst} or diamagnetic magnetization \cite{YBCOMag} above $T_c$ . The temperature range of SCFs in iron-based superconductors seem not so wide \cite{SCFLaOFeAs,SCFBaFeCoAs,SCFFeSeTe,SCFBaFeAsP,SCFLiFeAs,ThermoReview} except in CsFe$_2$As$_2$ \cite{SCFCsFeAs}. Recently, it is suggested that SCFs may be very strong in FeSe single crystals because of the vicinity to the Bose-Einstein condensate (BEC) and Bardeen-Cooper-Schrieffer (BCS) crossover region of the material \cite{MatsudaPNAS,MatsudaNC}. This recommends a picture of preformed Cooper pairs with phase incoherence far above $T_c$ in FeSe.

In this work, we carry out careful measurements using multiple tools on FeSe single crystals, allowing us to get further information about the SCF effect above $T_c$. We observe BCS mean-field like SCFs in a rather narrow temperature range in this material, which is similar to the situation in other iron-based superconductors, such as optimally doped Ba$_{1-x}$K$_{x}$Fe$_2$As$_2$ (BaK122). This may be attributed to the very small Ginzburg number in this family of superconductors. However, when counting the conduction electrons in the coherence volume, we indeed find that the density of the Cooper pairs in real space is much diluted, near the BCS-BEC crossover region.

\section{Experimental techniques}
High-quality FeSe single crystals were grown by chemical vapor transport method with the eutectic mix of KCl and AlCl$_3$ as the transport agent \cite{AEBohmer2013PRB}. Fe$_{1.04}$Se polycrystals were grown as the starting materials by solid state reaction. The mixture of Fe$_{1.04}$Se, KCl and AlCl$_3$ (molar ratio 1:2:4) was put into the bottom of a quartz ampoule, and the quartz ampoule was sealed under vacuum. The quartz ampoule was then placed into a horizontal tube furnace and heated up to 430 $^\circ$C. After keeping temperature for 30 hours to melt the transport agent, we create a temperature gradient by lowing the temperature of the end of the ampoule without reactant down to 370 $^\circ$C. We then kept sintering the sample with this temperature gradient for 6 weeks, and FeSe single crystals with tetragonal structure were obtained at the colder end of the ampoule.

The resistivity measurements were carried out in a Quantum Design Physical Property Measurement System (PPMS) by conventional four-probe method at different fields. The magnetization was measured by a Quantum Design SQUID-VSM, and the magnetic field was always applied in parallel to the $c$-axis of the sample in the measurements. We measured the specific heat by using a thermal-relaxation method which was an option of PPMS. The Nernst effect was measured by the one-heater-two-thermometer technique on a home-made setup attached to PPMS; a temperature gradient along the length direction was established by a heater and the temperature difference was measured by a pair of type-E thermocouples. The transverse voltage of the Nernst signal is measured at a magnetic field perpendicular to the thermal current in the $ab$-plane, and the data were taken with both positive and negative fields to reduce the interference of thermopower signal which is field symmetric.

\section{RESULTS}

\subsection{Magnetization and analysis near $T_c$}

\begin{figure}
\includegraphics[width=8cm]{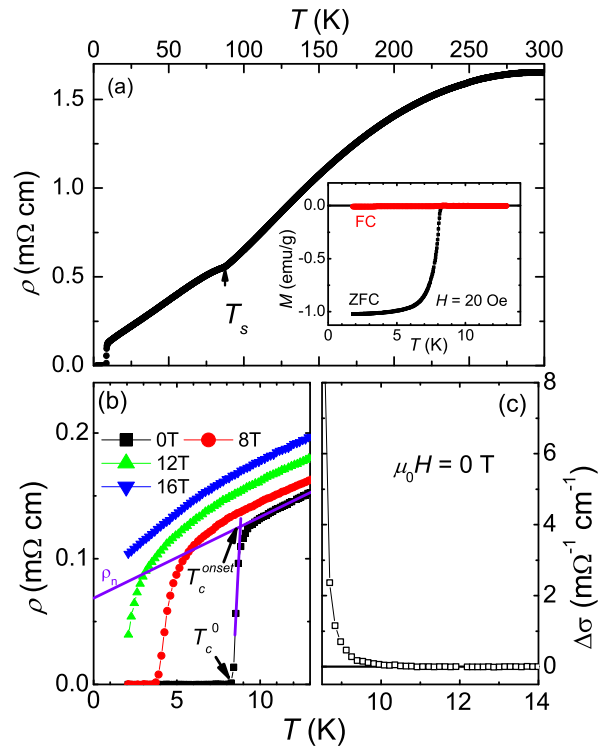}
\caption{(Color online) (a) Temperature dependence of resistivity at zero magnetic field. The inset shows the temperature dependence of magnetization measured in ZFC and FC modes at 20 Oe. (b) Temperature dependent resistivity measured at magnetic fields of 0, 8, 12 and 16 T. (c) Temperature dependence of excess conductivity at 0 T calculated by the data in (b).
} \label{fig1}
\end{figure}

The of Fig.~\ref{fig1}(a) shows the temperature dependence of zero-field-cooled (ZFC) and field-cooled (FC) magnetization measured at 20 Oe. The bulk superconducting transition temperature $T_c\approx8.7$ K is determined from the onset of the magnetization transition from the enlarged view near $T_c$. The temperature dependence of resistivity $\rho(T)$ at zero magnetic field is shown in Fig.~\ref{fig1}(a). A kink can be clearly identified at $T_s\approx 87.4$ K which is caused by the structural transition from tetragonal to orthorhombic phase, or the establishment of the nematic state. The superconducting transition temperature $T_c^{onset}$ is about 8.8 K determined in Fig.~\ref{fig1}(b) using the usual crossing method, and zero resistivity occurs at $T_c^{0} = 8.3$ K with a transition width of about 0.5 K. The residual resistivity ratio $RRR=\rho(300\ \mathrm{K})/\rho (0\ \mathrm{K})$ is about 24. The zero temperature resistivity $\rho (0\ \mathrm{K})$ is determined through a linear fit to the low-temperature data between 10 and 14 K. The resistivity curves measured at finite magnetic fields exhibit a clear enhancement compared to that measured at zero field, and the magnetoresistance defined as $MR =\left[\rho(16\ \mathrm{T})-\rho(0\ \mathrm{T})\right]/\rho(0\ \mathrm{T})$ is about 43\% at 10 K. This $MR$ value is much smaller than the ones from previous work \cite{MatsudaPNAS,MatsudaNC}. However it should be noted that the huge magnetoresistance in previous report \cite{MatsudaPNAS,MatsudaNC} is followed by an insulating-like upturn in the low-temperature region. This giant $MR$ observed in other work suggests that some exotic physics may be involved here, for example it is more or less similar to the pressure induced resistance upturn associated with some magnetic ordering \cite{ChengJGPressure,CanfieldPressure}, or due to the quantum oscillation effect and the density of states shows a significant decreasing under the magnetic field. The excess conductivity is defined as $\Delta\sigma=1/\rho(T)-1/\rho_n(T)$ with $\rho_n$ the linear extrapolation of the normal-state resistivity, which may be caused by the existence of residual Cooper pairs above the bulk $T_c$. A large excess conductivity is usually regarded as the mark of SCFs. The calculated excess conductivity from Fig.~\ref{fig1}(b) is shown in Fig.~\ref{fig1}(c), and the upper limit temperature of the SCF region seems to be less than 11 K. It should be noted that temperature dependent normal-state resistivity of FeSe is more complex than a straight line, and the upper temperature limit of the excess conductivity is dependent on the fitting temperature range. It is very difficult to define the SCF region from resistive measurement, thus we measure magnetization in the temperature region near $T_c$ to check how strong the SCFs are in the material.

\begin{figure}
\includegraphics[width=8cm]{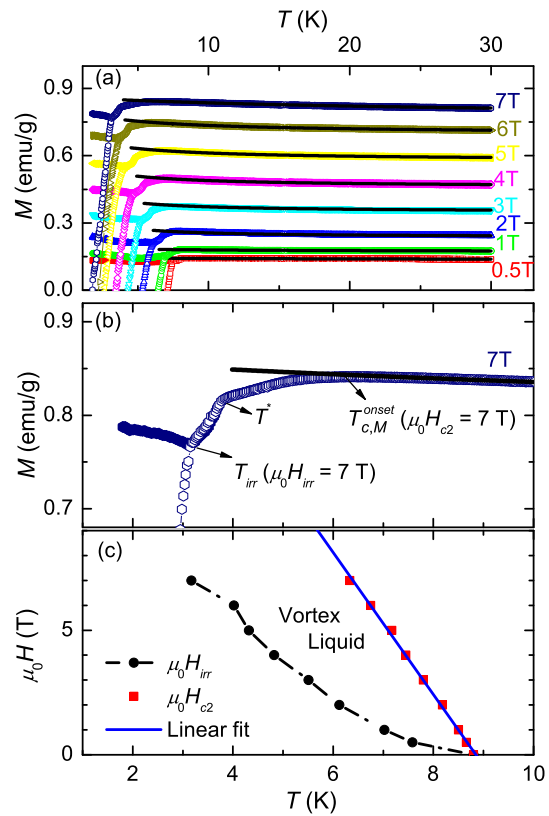}
\caption{(Color online) (a) Temperature dependence of mass magnetization measured at different fields. The open and solid symbols are the data measured with ZFC and FC modes, respectively. The black solid lines are the fitting curves by using Curie-Weiss law. (b) The enlarged view of the magnetization data at 7 T and the definitions of characteristic temperatures at characteristic magnetic fields. (c) The $\mu_0H$-$T$ vortex phase diagram of FeSe from the magnetization measurements. The solid squares represent the experimental data of $\mu_0H_{c2}(T)$. The solid blue line is a linear fit to the upper critical field $\mu_0H_{c2}(T)$ near $T_c$.
} \label{fig2}
\end{figure}

Fig.~\ref{fig2}(a) depicts the temperature-dependent magnetization curves measured with ZFC and FC modes at strong magnetic fields. A large positive magnetization background is observed for each field in the normal state, which may smear the weak diamagnetization arising from the SCFs. The positive magnetization background in the normal state originates from paramagnetic impurities and Pauli paramagnetism of the conduction electrons, which is supported by the monotonically increase of $M$ with decreasing temperature above $T_c$.  In order to get the net contribution from possible SCFs, we fit the ZFC magnetization data in the normal state by the Curie-Weiss law as,
\begin{equation}
M=M_0+C_0/(T+T_{\theta})\label{eq6}
\end{equation}
where $M_0$, $C_0$ and $T_{\theta}$ are the fitting parameters. The first term $M_0$ in Eq.~\ref{eq6} represents mainly the Pauli paramagnetization part contributed by conduction electrons. The second term, $C_0/(T+T_{\theta})$ is contributed by the magnetic moments in the sample, where $T_{\theta}$ denotes the Curie temperature. The fitting parameters at various fields are shown in Table.~\ref{table2}.

\begin{table}
\caption{The parameters derived from fitting to the $M$-$T$ data in Fig.~\ref{fig2}(a) by using Curie-Weiss law as Eq.~\ref{eq6}.}
\begin{ruledtabular}
\begin{tabular}{cccc}
$\mu_0H$ (T)&$M_0$ (emu/g) & $C_0$ (emu$\cdot$K/g)  &$T_\theta$ (K)  \\
\hline
0.5 &  0.13&  0.11 & 17.17 \\
1 & 0.17 &0.15   &8.34   \\
2 &  0.23&0.19   &0.10   \\
3 &  0.35&0.25   &1.21   \\
4 &  0.46&0.35   &2.48   \\
5 &  0.58&0.45   &3.47   \\
6 &  0.70&0.55   &4.91   \\
7 &  0.77&2.20   &24.67   \\
\end{tabular}
\end{ruledtabular}\label{table2}
\end{table}

The fitting curves are shown by solid lines in Fig.~\ref{fig2}(a). One can see that the Curie-Weiss law describes the experimental data quite well in the normal state. The deviation point between the magnetization curve and the fitting curve is defined as an onset temperature $T_{c,M}^{onset}$ at which the upper critical field $\mu_0H_{c2}$ equals to the applied magnetic field. A particular case for 7 T is shown as an enlarged view in Fig.~\ref{fig2}(b). Although there is an uncertainty in defining the very onset temperature $T_{c,M}^{onset}$, one can see that the allowed region of temperature is quite narrow (about $\pm 1$ K). We can also determine the irreversible temperature $T_{irr}$ by taking the separation point of ZFC and FC magnetization curves, and at that temperature $\mu_0H_{irr}=7$ T in Fig.~\ref{fig2}(b). It should be noted that there is a kink in the magnetization curves between $T_{c,M}^{onset}$ and $T_{irr}$, which is marked as $T^*$ and is also field dependent. The characteristic temperature $T^*$ may be caused by some unknown vortex phase transition in the vortex liquid region, and it needs further experimental investigation. We leave the discussion on $T^*$ to a separate study. The $\mu_0H$-$T$ phase diagram of FeSe shown in Fig.~\ref{fig2}(c) displays the temperature-dependent behavior of $\mu_0H_{c2}$ and $\mu_0H_{irr}$. One can find that the vortex liquid region in the phase diagram is not so wide. The almost linear temperature-dependent $\mu_0H_{c2}$ near $T_c$ gives rise to a slope $d\mu_0H_{c2}/dT=-2.85$ T/K near $T_c$, and the calculated $\mu_0H_{c2}(0)$ at zero temperature by the Werthamer-Helfand-Hohenberg (WHH) formula \cite{WHH}, i.e., $\mu_0H_{c2}(0)=0.69T_c|d\mu_0H_{c2}/dT|_{T_c}$ is about 15.7 T. This is consistent with a previous report \cite{MatsudaPNAS}. The coherence length at zero temperature $\xi_0$ can be obtained according to the formula $\mu_0H_{c2}(0)=\Phi_0/(2\pi\xi_0^2)$ with $\Phi_0$ the magnetic flux quantum. Considering $\xi_0=\hbar v_F/(\pi\Delta_0)$ approximately in a plain $s$-wave superconductor ($v_F$ the Fermi wave velocity and $\Delta_0$ the gap at $T=0$) and Fermi energy $E_F=m^*v_F^2/2$, the effective mass can be estimated by
\begin{equation}
m^*=\frac{4\hbar^2}{\pi\Phi_0}\frac{\mu_0H_{c2}(0)E_F}{\Delta_0^2}.
\end{equation}
The gap maximum from the STS measurements \cite{DavisQPI} is 2.3 meV for the hole pocket around $\Gamma$ point and 1.5 meV for the electron pocket around M point. In addition with $E_F$ for different bands \cite{QuanOs1}, we can obtain $m^*=2.5m_e$ for $\delta$ branch of the hole pocket, and $m^*=3.0m_e$ for $\gamma$ branch of the electron pocket, with $m_e$ the free electron mass. The estimated values are comparable with $4.3m_e$ and $7.2m_e$ for these selected branches of the Fermi cylinders with maximal cross sections from Shubnikov-de Haas oscillation measurements \cite{QuanOs1}. The acceptable difference of effective mass from two different methods may be related to the characters of multiband and highly anisotropic superconducting gaps in FeSe. We emphasize that, from our magnetization data and analysis, the SCF region is quite limited. For example, at a field of 7 T, the upper limit temperature for SCFs is only about 7 K.

\begin{figure}
\includegraphics[width=8cm]{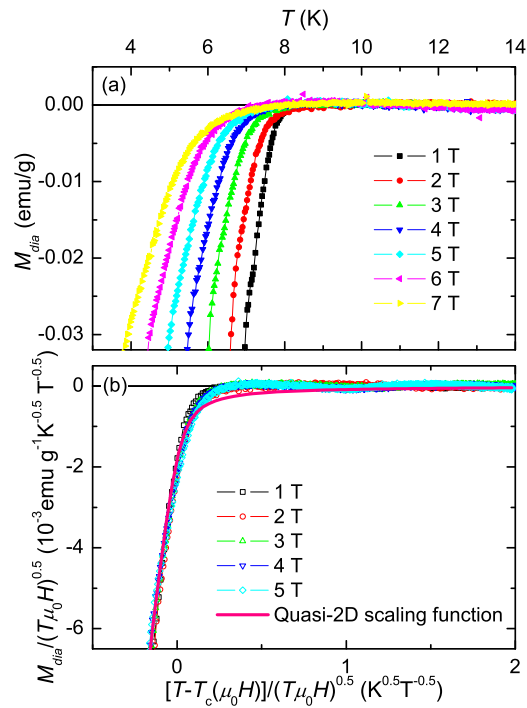}
\caption{(Color online) (a) Temperature dependence of diamagnetic magnetization $M_{dia}$ at different fields obtained by subtracting the fitted background by Eq.~\ref{eq6} from the experimental data. (b) Scaling curves from (a) by using the Ginzburg-Landau fluctuation theory for a quasi-2D system. The solid line is the expected scaling function of the quasi-2D scaling theory.} \label{fig3}
\end{figure}

In order to investigate the SCFs in FeSe, Ginzburg-Landau fluctuation theory is used to analyze the diamagnetic signal near $T_c$. Since the normal state has a background of magnetization, we need to subtract this part from the total signal. The magnetization $M_{dia}$ arising from superconductivity (including SCFs) shown in Fig.~\ref{fig3}(a) are obtained by subtracting the fitted paramagnetic background (Curie-Weiss term) from the measured magnetization. One can see that $M_{dia}$ approaches to zero just above $T_c$, which indicates that SCFs are very weak. Usually a common crossing point or a small crossing area appears on the set of $M$-$T$ curves at different magnetic fields in a superconductor with strong SCFs, which can be well described by quasi-two-dimensional (quasi-2D) or three-dimensional (3D) lowest Landau level (LLL) scaling formula based on Ginzburg-Landau (GL) theory \cite{Tesanovic}. This has been well studied in cuprate superconductors \cite{GaoHPRB,WenHHPRL1999}. However, we find that the $M_{dia}(T)$ curves in our present sample separate from each other, showing no crossing point or area expected by the GL-LLL scaling theory. Although this situation suggests that the scaling theory fails in this system, we still try to scale the $M_{dia}$-$T$ curves by following the scaling law to obtain further comprehension. By using a non-perturbative approach to the GL free energy function for a quasi-2D system, the $M(T)$ curves can be scaled by
\begin{equation}
M/(T\mu_0H)^{0.5}=Cf\left\{\frac{A[T-T_c(\mu_0H)]}{\sqrt{T\mu_0H}}\right\},\label{eq1}
\end{equation}
where
\begin{equation}
f(x)=x-\sqrt{x^2+2},\label{eq2}
\end{equation}
and the field-dependent $T_c$ is expressed as
\begin{equation}
T_c(\mu_0H)=T_{c0}-\mu_0H\left(\frac{d\mu_0H_{c2}}{dT}\right)^{-1}.\label{eqTc}
\end{equation}
Here $T_{c0}$, $d\mu_0H_{c2}/dT$, $A$, and $C$ in Eq.~\ref{eq1} and \ref{eqTc} are the scaling parameters. The parameter $A$ is dependent on the GL parameter $\kappa$ and $|d\mu_0H_{c2}/dT|_{T_c}$, and $C$ is inversely proportional to $\kappa$. In addition both of $A$ and $C$ are independent of $H$ or $T$. The quasi-2D scaling curves at different fields in FeSe are shown in Fig.~\ref{fig3}(b), where $M_{dia}/(T\mu_0H)^{0.5}$ is scaled as a function of $[T-T_c(\mu_0H)]/(T\mu_0H)^{0.5}$. Surprisingly, the quasi-2D fluctuation scaling law works well on the data measured from 1 T to 5 T although there is no crossing point for the $M_{dia}$-$T$ curves. The data at 6 and 7 T show clear deviation from the scaling (not shown here). The parameters $T_{c0}$ and $d\mu_0H_{c2}/dT$ obtained from the scaling process are $8.12$ K and $-2.13$ T/K respectively, which are comparable to the values obtained from the resistive and magnetization measurements. We have also tried the 3D GL fluctuation scaling, which does not work on the data. The absence of crossing point in $M_{dia}$-$T$ curves and the nice quasi-2D fluctuation scaling behavior seem contradictory in this material. However, as shown in Fig.~\ref{fig3}(b), the scaled curves deviate from the scaling function of Eq.~\ref{eq2}. The difference between the scaling curves and the required scaling function is more pronounced in the fluctuation region, and the scaled curves show much narrower SCF region compared with the quasi-2D scaling function shown by the pink solid line here. Therefore, the scaling behavior in FeSe can not be described by the GL-LLL scaling theory, which is consistent with the absence of the crossing point in the diamagnetic magnetization curves.

\subsection{BCS mean-field like transition detected by specific heat measurement}

\begin{figure}                                                                                                                                                                                                               \includegraphics[width=8cm]{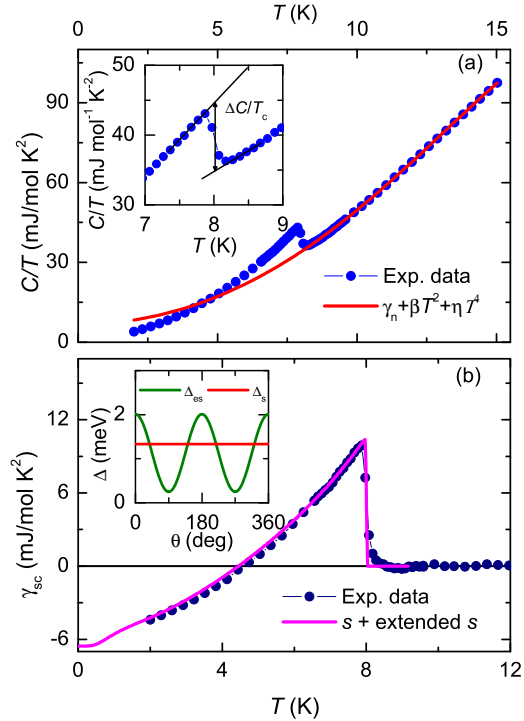}                                                                                                                                                                                        \caption{(Color online) (a) Temperature-dependent specific heat measured at 0 T. The red solid line is the normal state fitting curve. The inset shows the enlarged view of the specific heat jump near $T_c$, and $\Delta C/T_c$ is estimated by entropy conservation near $T_c$.
(b) Temperature dependence of superconducting electronic specific heat. The transition temperature $T_c^{SH}$ from specific heat jump estimated by entropy conservation as show in (b) or the inset in (a) is about 8 K. The blue line represents the fitting curve using the BCS formula. The inset shows the angle-dependent gap functions used in the fitting.
} \label{fig4}                                                                                                                                                                                                               \end{figure}

Specific heat is a useful tool to detect SCFs near $T_c$. Figure~\ref{fig4}(a) shows the specific heat as a function of temperature at 0 T. As we can see from the enlarged view in the inset of Fig.~\ref{fig4}(a), there is a sharp jump in $\Delta C/T_c$ near the superconducting transition. The specific heat jump estimated by entropy conservation yields the ratio $\Delta C/\gamma_n T_c=1.46$, which is close to 1.43 predicted by BCS theory in the weak coupling regime. In order to obtain the superconducting electronic specific heat, we fit the data above $T_c$ by $C_n/T =\gamma_n+\beta T^2 +\eta T^4$, where $\gamma_n$ is the normal-state electronic part, and $\beta T^2 +\eta T^4$ is the phonon contribution by Debye model in low-temperature region. The fitting result is shown by the red solid line in Fig.~\ref{fig4}(a); the parameters obtained from the fit are $\gamma_n =6.7$ mJ$\cdot$mol$^{-1}\cdot$K$^{-2}$, $\beta=0.41$ mJ$\cdot$mol$^{-1}\cdot$K$^{-4}$ and $\eta=3.4\times10^{-4}$ mJ$\cdot$mol$^{-1}\cdot$K$^{-6}$. Temperature dependence of superconducting electronic specific heat obtained by the equation $\gamma_{sc}=C_{sc}/T = (C- C_n)/T$ is shown in Fig.\ref{fig4}(b), and then we fitted the data by BCS formula. The superconducting electronic specific heat $\gamma_{sc}$ for an anisotropic superconducting gap can be expressed as
\begin{eqnarray}
\gamma_{sc}=\frac{4N(E_F)}{k_BT^{3}}\int_{0}^{+\infty}\int_0^{2\pi}\frac{e^{\zeta/k_BT}}{(1+e^{\zeta/k_BT})^{2}}\nonumber\\
\nonumber\\
\cdot\left[\varepsilon^{2}+\Delta^{2}(\theta,T)-\frac{T}{2}\frac{d\Delta^{2}(\theta,T)}{dT}\right]\,d\theta\,d\varepsilon,
\end{eqnarray}
where $\zeta=\sqrt{\varepsilon^2+\Delta^2(T,\theta)}$. A linear combination of two components with different gaps, namely, $\gamma_{sc}=x\gamma_{sc1}(\Delta_s)+(1-x)\gamma_{sc2}(\Delta_{es})$, is used to describe the experimental data as previous report \cite{J.Y.LinPRB2011}. The gap functions we used are an $s$-wave $\Delta_s$ and an extended nodeless $s$-wave $\Delta_{es}=\Delta_{es}^0(1 +\alpha\cos2\theta)$. A set of fitting parameters we choose for the experimental data are $\Delta_s(0)=1.33$ meV, $\Delta_{es}^0(0)=1.13$ meV, $x=0.2$, and $\alpha=0.78$. The angle dependent gap functions used for the fitting are shown in the inset of Fig.~\ref{fig4}(b), and the fitting curve is shown as the solid curve in Fig.~\ref{fig4}(b) with a very sharp transition at $T_c^{SH}=8$ K. We should notice that it is difficult to obtain the precise gap structure by fitting the data in Fig.~\ref{fig4}(a) due to the lack of data measured at extremely low temperatures. The feature near 1 K on the fitting curve is lack of experimental data support. The gap symmetry is very sensitive to the low temperature data, and such will be presented in a separate publication \cite{ChenGY2017}. Here the fitting curve can be regarded as a guide line which satisfies the entropy conservation. Other gap functions \cite{J.Y.LinPRB2011,ChenGY2017} will have very little influence on the shape of specific jump of the fitting curve near $T_c$. One can clearly see that, the temperature range of SCFs illustrated by the extending tail of $\gamma_{sc}$ above $T_c^{SH}$ is very narrow, and the highest fluctuation temperature can only be extended to about 9 K. This value is quite close to the onset transition temperatures from the transport and magnetization measurements. Hence the specific heat data near the superconducting transition can be well described by BCS mean-field approach, which confirms the narrow fluctuation region in FeSe.

In the framework of BCS-Eliashberg mean-field theory, the formation and condensation of Cooper pairs take place at the same temperature $T_c$ \cite{Emery1995Nature}. However, if phase fluctuations are too strong, the Cooper pairs may preform at a temperature $T^*$ higher than the Cooper-pair condensation temperature $T_c$, hence there will be a wide temperature range of residual superconductivity between $T_c$ and $T^*$ \cite{Wen2009PRL}. The difference between the two theories mentioned above can be easily distinguished by the shape of the specific heat curve near $T_c$ \cite{Alain1999PhysicaC}. In the ideal BCS mean-field case, a second-order phase transition takes place at $T_c$, which leads to a very sharp jump in specific heat. In the system with moderate SCFs, a $\lambda$-shaped transition of specific heat coefficient will be observed, i.e., the high-temperature end point $T^*$ is a bit away from $T_c$. If the SCFs are strong enough, temperature difference between the real-space pairing and the bulk superconductivity governed by BEC is very large. The specific heat decreases smoothly from $T_c$ to $T^*$, and the shape of the specific heat data is similar to the one in the $\lambda$-transition of $^4$He from a normal fluid to a superfluid near 2.17 K \cite{lambdaHe4}. As we can see from the data in FeSe, the specific heat data shows only a very small tail above $T_c^{SH}$, which can be described quite well by the BCS mean-field transition and is very different from the picture of BEC.

\subsection{Narrow SCF region verified by Nernst effect}

\begin{figure}
\includegraphics[width=8cm]{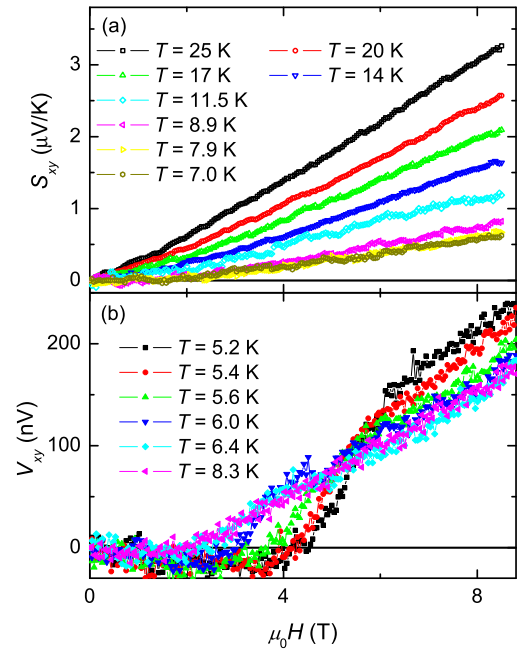}
\caption{(Color online) (a) Field-dependent Nernst signal $S_{xy}$ at different temperatures above 7 K. (b) Field dependence of Nernst transverse voltage $V_{xy}$ at different temperatures with tiny exchange helium gas to lower down the base temperature of the sample.
} \label{fig5}
\end{figure}

The vortices in the flux-flow state of a superconductor carry entropy within and nearby the vortex core, and they will move from the hot end to the cold one of the sample with temperature gradient. Transverse Nernst signal is sensitive to the vortex motion when the magnetic field is applied perpendicular to the thermal current. The field (or temperature) dependent Nernst signal $S_{xy}=E_y/\partial_xT$ at some fixed temperature (or field) dominated by vortex motion is usually hump-like. It means that $S_{xy}$ starts from zero when the vortices start moving, and reaches its maximum at some magnetic field (or temperature). $S_{xy}$ then decreases with increasing magnetic field (or temperature), and finally disappears in the normal state. Thus the Nernst signal in the flux flow region and at a fixed temperature can be written as $S_{xy}\propto H(1-H/H_{c2})$ \cite{Maki}. The Nernst signal may have a small tail above bulk $T_c$ or $H_{c2}$ because of the SCFs \cite{BehniaReview}. However, in cuprates, it was found that the Nernst signal has a hump-like field dependence in a very wide temperature range above $T_c$, which is regarded as strong SCFs \cite{CuNernstReview}.

We measure the Nernst effect to detect the SCFs in FeSe. The field-dependent curves of Nernst signal $S_{xy}$ at different temperatures above 7 K are presented in Fig.~\ref{fig5}(a). One can find that $S_{xy}$ is almost proportional to the magnetic field with a very weak contribution from a quadratic and even higher-order field terms. In a single-band metal, the Nernst coefficient $\nu_{xy}=S_{xy}/\mu_0H$ is very small because of the cancellation effect between the thermal and the coulomb contributions \cite{BehniaReview}. However, the value of $\nu_{xy}$ in FeSe seems very large even if compared with the typical multiband material NbSe$_2$ \cite{BehniaNbSe2}. The almost linear field-dependent Nernst signal with a slight positive curvature above $T_c$ is obviously from the normal-state properties, such as the bi-parity motion of electron and hole-like charge carriers. Even when the temperature decreases just below $T_c$, there is no obvious hump feature or negative second derivative originated from the vortex motion. The Nernst signal measurements are usually carried out at high vacuum to measure the exact value of the temperature gradient. However the lowest temperature of the sample with heating power is only about 6.5 K in our home-made measurement system in PPMS. We add some exchange helium gas to the measurement chamber to lower down the temperature when measuring the Nernst signal at lower temperatures. In this case, we cannot measure the exact temperature gradient. We plot the measured Nernst voltage $V_{xy}$ in Fig.~\ref{fig5}(b), and the relatively larger voltage noise is caused by temperature fluctuations from the exchange gas. The same heating power is applied for $V_{xy}$ measurements at temperatures from 5.2 to 8.3 K with an overlapped temperature range as the exact $S_{xy}$ measurements. Since the temperature range is only 3 K in $V_{xy}$ measurements, we can regard the temperature gradient as a constant in these measurements.

\begin{figure}
\includegraphics[width=8cm]{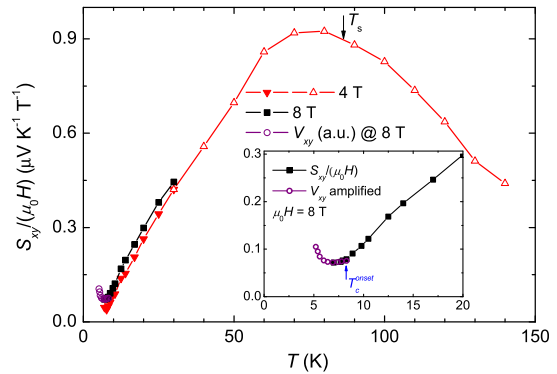}
\caption{(Color online) Temperature dependence of the Nernst coefficient at different fields. The filled symbols represent the data obtained by the field-dependent Nernst signal. The open triangles are Nernst coefficients measured at 4 T and different temperatures. Nernst voltage values measured with some exchange helium gas are shown as open circles with the arbitrary unit, i.e., the $V_{xy}$ values are multiplied by a necessary factor to make the data have a smooth connection to the Nernst signal $S_{xy}$ measured at the same field and temperature. The inset shows the enlarged view of the Nernst coefficient at low temperatures and 8 T.
} \label{fig6}
\end{figure}

The temperature dependent Nernst coefficient $\nu_{xy}$ and amplified $V_{xy}$ measured with exchange gas are shown in Fig.~\ref{fig6}. One can see that the Nernst coefficient has a huge hump with maximum value as large as 0.9 $\mu$V$\cdot$K$^{-1}\cdot$T$^{-1}$ below and near the structural transition temperature $T_s$. Similar data have been measured on the parent compounds of iron pnictides with both structural transition and spin-density-wave (SDW) transition \cite{NernstLaFeAsO,NernstCaFe2As2,NernstEuFe2As2}. Such a large Nernst signal with a hump like temperature dependence are quite often observed below the SDW transition temperature, so it is suggested that the SDW order or SDW fluctuation may enhance the Nernst coefficient \cite{ThermoReview}. However, there is only structural transition in FeSe without any magnetic order in the sample at zero pressure \cite{YuWQPRL}. One of the explanations for the huge Nernst coefficient is some possible spin fluctuations in FeSe which is too weak to be observed on the sample. It should be noted that the Hall coefficient in all these materials with huge Nernst peaks seems very small and even has a sign change with increasing temperature. Therefore, another possibility of the huge Nernst signal may come from the almost balanced hole and electron pockets \cite{BehniaNbSe2}; the drastic change of the Nernst signal near $T_s$ may be induced by the change of Fermi surface near the structural transition.

The Nernst coefficient exhibits a small enhancement with decreasing temperature below $T_c$. At the lowest measured temperature (5.2 K), the Nernst signal is zero below 4 T, and then it ramps up at higher fields, as shown in Fig.~\ref{fig4}(b). At about 6 to 7 T, the curve shows a negative curvature and merges into the background of the normal state. Similar situation occurs for the data at temperatures up to 6.4 K, while the threshold of magnetic field for flux flow now becomes much smaller. This phenomenon is regarded as the typical feature of vortex flow on top of a large background signal, although we did not obtain the peak of the Nernst signal in the whole region. A peak associated with flux flow is absent, because the Nernst signal arising from vortex motion is much weaker than the normal-state background. Hence the Nernst coefficient shown in Fig.~\ref{fig6} from the vortices become negligible near and above $T_c$, which is consistent with the conclusion from field dependent Nernst signal. The Nernst effect measured in FeSe is similar to the situation in Fe$_{1+y}$Te$_{0.6}$Se$_{0.4}$ where SCFs are very weak and the amplitude of the peak of the Nernst signal from the vortices is also very small \cite{SCFFeSeTe}. In electron doped cuprates, one also see very similar situation, i.e., the Nernst signal with a very small peak structure appears on top of a large background due to bi-parity contributions \cite{PengchengLi2007PRB}. One explanation for this very small Nernst signal is that the vortices in FeSe or Te doped FeSe may carry very little entropy, together with the fact that the vortex liquid region is very small. In any case, we can conclude that no vortex motion induced Nernst signal is observed above $T_c$.

\subsection{Revisit the calculation of the Ginzburg number}
SCFs in a superconductor can be described approximately by the GL theory, and the magnitude of SCFs can be characterized by the Ginzburg number \cite{BlatterReview}
\begin{equation}
Gi=\left(k_BT_c/\mu_0H_c^2\epsilon\xi^3\right)^2/2.\label{eq4}
\end{equation}
Here the thermodynamic critical field $\mu_0H_c=\mu_0H_{c2}/\left(\sqrt{2}\kappa\right)$. The upper critical field $\mu_0H_{c2}$ is obtained by extrapolating the linear part of $\mu_0H_{c2}(T)$ near $T_c$ and calculated \cite{LobbPRB1987} by using the WHH formula $\mu_0H_{c2}(0)=-0.69T_c[d\mu_0H_{c2}(T)/dT]_{T_c}$, instead of using the measured value of $\mu_0H_{c2}(0)$. The anisotropy parameter $\epsilon\equiv\sqrt{m^*_{ab}/m^*_c}=H_{c2}^{\parallel c}/H_{c2}^{\parallel ab}$ is usually smaller than 1. Considering the relationship between $H_{c2}$ and $\xi$, i.e., $\xi=\sqrt{\phi_0/2\pi\mu_0H_{c2}}$, the Ginzburg number $Gi$ calculated in the SI unit reads
\begin{equation}
Gi=1.7\times10^{-11}T_c^2\kappa^4/\left(\mu_0H_{c2}\epsilon^2\right).\label{eq5}
\end{equation}
Here the coefficient and all the parameters are in SI units. It should be noted that the original calculation in Ref.~48 have some mistakes in using the value of $\mu_0$ in cgs units, therefore the exact value of the frequently discussed $Gi$ should be multiplied by a factor of $(4\pi)^2$. Because of this error, in high-$T_c$ cuprate superconductors, the mistakenly used Ginzburg number is supposed to be $Gi\approx10^{-3}\sim10^{-1}$, while the value is extremely small ($Gi\approx10^{-8}\sim10^{-6}$) in conventional superconductors \cite{BlatterReview}. However, as just mentioned, the real values of $Gi$ should be multiplied by $(4\pi)^2$, which enhances $Gi$ two orders of magnitude larger than the previously widely used ones. We want to emphasize that, in order to have a meaningful discussion on the SCFs and their temperature range, $Gi$ should be much smaller than 1, but the calculated value of $Gi$ in some cuprates, like Bi2212, may be greater than 1. Although such a large value of $Gi$ is not meaningful, it may just suggest strong fluctuations in the sample. In addition, the temperature range of SCFs is determined simply by the Ginzburg number as $GiT_c$ where $Gi$ seems to be too small in the original calculation \cite{LobbPRB1987}. For example, for YBa$_2$Cu$_3$O$_{7-\delta}$ (YBCO), the calculated $Gi$ using original formula is 0.00127, and the SCF region will be only 0.11 K. This is unreasonable. Therefore a factor of $(4\pi)^2$ should be multiplied to the original formula, or the correct form of calculating $Gi$ is Eq.~\ref{eq5}.

\begin{table}
\caption{Characteristic superconducting parameters of different superconductors.}
\begin{ruledtabular}
\begin{tabular}{ccccccc}
             & Bi2212  & YBCO & MgB$_2$ & BaK122 & FeSe \\
\hline
$T_c$ (K)    & 95  & 91                & 39                    & 38                    & 8.2 \\
$\mu_0H_{c2}$ (T)  & 177 & 180                & 4                     & 180                    & 16   \\
$\kappa$      &  115    & 62                & 21                   & 80                    & 72   \\
$\epsilon$  & 0.02   & 0.24               & 0.5                     & 0.5                   & 0.55   \\
$Gi$       & 380   & 2$\times$10$^{-1}$  & 5$\times$10$^{-3}$  & 2.2$\times$10$^{-2}$    & 6.4$\times$10$^{-3}$   \\
\hline
$T_c/T_F$   & 0.035 & 0.011     & 0.007         &     0.17                & 0.2   \\
$k_F\xi$   & - &  - &   23   &                      4 &  3 \\
$V_{coh}n_{pair}$  &  1  & 109             & 1.5$\times$10$^{5}$                     &  122                & 31   \\
\end{tabular}
\end{ruledtabular}\label{table1}
\end{table}

In the following, we roughly estimate on the values of $Gi$ and the SCF regions for different superconductors. We present the typical superconducting parameters and calculated $Gi$ for Bi$_2$Sr$_2$CaCu$_2$O$_8$ (Bi2212) \cite{Qiangli1993PRB,EFSC,SCPara}, YBCO \cite{Orlando,Nguyen,Zimmermann,EFSC,SCPara}, MgB$_2$ \cite{MgB2Review,Caplin,EFSC,SCPara}, Ba$_{0.6}$K$_{0.4}$Fe$_2$As$_2$ (BaK122) \cite{H.Yang,C.Ren,Tarantini,ARPESBa122,SCPara}, and FeSe \cite{MatsudaPNAS,Terashima,Hafiez} in Table~\ref{table1}. Here $\mu_0H_{c2}$ is determined from the value of $d\mu_0H_{c2}/dT$ near $T_c$. The calculated $Gi$ for BaK122 or FeSe are much smaller than those in Bi2212 or YBCO, but comparable with the value in MgB$_2$. Meanwhile it is claimed that another calculation method for $Gi$ in a 2D system is to determine the ratio $T_c/T_F$, and it is of the order of $10^{-1}$ in BEC limit and $10^{-5}\sim10^{-4}$ for BCS superconductors \cite{FluctuationSC}. The calculated values of $T_c/T_F$ from $E_F$ in 2D cuprate superconductors Bi2212 and YBCO are very small because of the large Fermi energy; the value of FeSe is comparable with the one in BaK122. This seems in conflict with the situation that Bi2212 is a typical superconductor with very strong SCFs, and people even argue that the superconductivity in Bi2212 is governed by a BEC-like transition \cite{Alain1999PhysicaC}. Actually the strong SCFs in Bi2212 are more likely to be driven by the strong anisotropy (or small $\epsilon$). Therefore the ratio $T_c/T_F$ may not be an appropriate parameter to determine the fluctuation behavior of a superconductor. In addition, for a 3D system, $Gi$ is determined \cite{FluctuationSC} by $80(T_c/T_F)^4$, the value of $Gi$ is further lowered down. For iron-based superconductors, such as FeSe and BaK122, it is closer to the 3D case, so a simple estimation of $Gi\approx T_c/T_F$ may cause problems. Furthermore, for a system with multi-bands, if some bands have large Fermi energies, while others have very small values, the superfluid coming from the band with large Fermi energy may stabilize the condensation and suppress the SCFs. Thus, a correct way to estimate SCFs is to use Eq.~\ref{eq5}. These arguments may answer the question why SCFs in bulk FeSe are not strong.

\section{Discussion and counting on the overlapped Cooper pairs}

In the BCS theoretical picture, the Cooper pairing and condensation occur simultaneously. In this framework, it is meaningless to describe a Cooper pair in real space. One can only say that many electrons form a highly entangled paired state. The basic reason for that is the strong overlap between Cooper pairs. Therefore in this model, tens of thousands of Cooper pairs are overlapping each other. The product of the Fermi vector $k_F$ and coherence length $\xi$, is a very good quantity to estimate how strong the overlap is. The quantity $k_F\xi$ tells roughly how much conduction electrons or Cooper pairs in one coherence length. Thus this parameter is also used to define the crossover from BCS to BEC \cite{F.Pistolesi} when $k_F\xi\approx1$. $k_F\xi\geq2\pi$ corresponds to BCS-like superconductivity, while $k_F\xi\leq1/\pi$ corresponds to the BEC case. Taking the related quantities from experiment, we find that both FeSe and BaK122 have the value of $k_F\xi$ smaller than $2\pi$ according to the calculation of the chemical potential, so they may be near the crossover region from BCS to BEC but much closer to BCS \cite{F.Pistolesi}. To consider the situation in 3D case, the number of superconducting electrons in unit coherent volume $V_{coh}n_{pair}$ can also be used \cite{SrTiO3Marel} to determine the different situations between BCS and BEC. Here coherent volume $V_{coh}=4\pi\xi_{ab}^2\xi_{c}/3$ for an anisotropic superconductor, while $n_{pair}=n_s/2$ is the density of the superconducting electrons with opposite momentum which can be paired with the selected one. Following the above discussions, we expect $V_{coh}n_{pair}\gg1$ in the BCS case, while $V_{coh}n_{pair}\ll1$ for the BEC limit \cite{Uemura}. The evolution from BCS to BEC is accompanied with the significant reduction of $V_{coh}n_{pair}$. We replace $n_s$ with the charge carrier density of the normal state $n$ approximately, and the calculated $V_{coh}n_{pair}$ for different superconductors are also listed in Table~\ref{table1}. The charge carrier density in FeSe is much larger than that in Nb doped SrTiO$_3$ \cite{BehniaSrTiO3} which is supposed to be a system with the very dilute superfluid density, but the calculated $V_{coh}n_{pair}$ is even much smaller than that in Nb doped SrTiO$_3$. Therefore it is reasonable to argue that the Cooper pairs in the coherence volume $V_{coh}$ in FeSe is diluted. It should be noted that the charge carrier density in unit coherent volume is still one order of magnitude larger than 1, and is comparable to the value in BaK122. From this issue, FeSe is not very different from other iron-based superconductors, like BaK122. Although $V_{coh}n_{pair}$ in BaK122 is close to that in YBCO, the specific heat measurement \cite{Ba122SH} in BaK122 shows a clear BCS mean-field like transition, and the SCF region in BaK122 is indeed very narrow as revealed by many different experimental techniques. We can conclude that both FeSe and BaK122 are far away from the BCS-BEC crossover region.

Here we have observed a rather narrow region of SCFs in FeSe above $T_c$, which is different from previous reports \cite{MatsudaPNAS,MatsudaNC}. One main difference is in the $M$-$T$ data. Kasahara et. al. observed a very weak diamagnetic signal within a wide temperature range above the bulk $T_c$ from the magnetization curves under high magnetic fields, which seems absent in our data. There are low-temperature upturn behavior in our $M$-$T$ data instead of the constant positive background in a previous report \cite{MatsudaNC}. The measured upturn behavior, which can be well described by the Curie-Weiss law, is mainly caused by paramagnetic impurities such as interstitial iron impurities. One possible reason for the absence of strong SCFs in our samples is that these paramagnetic impurities may suppress SCFs by inducing pair breaking to the preformed Cooper pairs above $T_c$. However, because the effective range of such a single impurity is very short, which is within about 10 $\AA$ in diameter as measured in Fe(SeTe) \cite{FeImpurityPan}, and the distance between these interstitial Fe impurities is quite large, we think that such paramagnetic impurities are unlikely to act as pair breakers to suppress strong SCFs. The relatively large $RRR$ value also supports that the scattering from the diluted impurities is very weak. Furthermore, the very small residual specific heat coefficient measured in the superconducting state indicates that the pair breaking by impurities is very weak \cite{ChenGY2017}. Additionally, one may argue that the SCF signal is buried in the upturn behavior of magnetic susceptibility from the paramagnetic impurities above $T_c$. The paramagnetic impurities are inevitable in a sample, however it should be noted that the upturn of magnetic susceptibility in our sample is very weak and can be well described by the Curie-Weiss law. This suggests that SCFs, if exist, would contribute a negligible signal above $T_c$. Clearly, more data from samples of different groups are required to verify the SCFs from magnetization measurements. However, based on our specific heat and Nernst data and thoughtful analysis mentioned above, we conclude that the SCF region in FeSe above $T_c$ is quite narrow. This indicates that even for a superconducting system with diluted Cooper pairs in the coherence volume, the SCFs are still very limited and the superconducting transition is still governed by the BCS mean-field like transition.

\section{Conclusion}
In conclusion, our magnetization, specific heat and Nernst effect studies all point to very weak superconducting fluctuations and a narrow SCF region above $T_c$ in FeSe. A revised calculation of the Ginzburg number $Gi$ using the standard method shows that $Gi$ is rather small in bulk FeSe, and close to other iron-based superconductors, such as BaK122. This explains our observation of a very narrow SCF region in bulk FeSe. The number of Cooper pairs (about 31) in the coherent volume is two or three orders of magnitude lower than that of a typical conventional superconductor, however the superconducting transition in FeSe is still governed by a BCS mean-field like critical transition. Theoretically, it is highly desired to understand why such theory still holds for superconducting systems with the dilute density of Cooper pairs.

\begin{acknowledgments}
We thank Yuji Matsuda, Christofer Meingast, and Tetsuo Hanaguri for helpful discussions, and Yaomin Dai for his assistance in revising the manuscript. This work was supported by the National Key Research and Development Program of China (2016YFA0300401 and 2016YFA0401700), National Natural Science Foundation of China (11534005,11374144), and Natural Science Foundation of Jiangsu (BK20140015).
\end{acknowledgments}

$^*$ huanyang@nju.edu.cn

$^{\dag}$ hhwen@nju.edu.cn


\begin{thebibliography}{41}

\bibitem{Mazin} I. I. Mazin, Nature (London) {\bf464}, 183 (2010).

\bibitem{Kuroki} K. Kuroki, S. Onari, R. Arita, H. Usui, Y. Tanaka, H. Kontani, and H. Aoki, Phys. Rev. Lett. {\bf101}, 087004 (2008).

\bibitem{StewartReview} G. R. Stewart, Rev. Mod. Phys. {\bf83}, 1589 (2011).

\bibitem{WenReview} H. H. Wen and S. L. Li, Annu. Rev. Condens. Matter Phys. {\bf2}, 121 (2011).

\bibitem{WuMKPNAS} F. C. Hsu, J. Y. Luo, K. W. Yeh, T. K. Chen, T. W. Huang, P. M. Wu, Y. C. Lee, Y. L. Huang, Y. Y. Chu, D. C. Yan, and M. K. Wu, Proc. Natl. Acad. Sci. USA {\bf105}, 14262 (2008).

\bibitem{SongCLScience} C. L. Song, Y. L. Wang, P. Cheng, Y. P. Jiang, W. Li, T. Zhang, Z. Li, K. He, L. L. Wang, J. F. Jia, H. H. Hung, C. J. Wu, X. C. Ma, X. Chen, and Q. K. Xue, Science {\bf332}, 1410 (2011).

\bibitem{ARPES1} T. Shimojima, Y. Suzuki, T. Sonobe, A. Nakamura, M. Sakano, J. Omachi, K. Yoshioka, M. Kuwata-Gonokami, K. Ono, H. Kumigashira, A. E. B\"{o}hmer, F. Hardy, T. Wolf, C. Meingast, H. v. L\"{o}hneysen, H. Ikeda, and K. Ishizaka, Phys. Rev. B {\bf90}, 121111 (R) (2014).

\bibitem{ARPES2} K. Nakayama, Y. Miyata, G. N. Phan, T. Sato, Y. Tanabe, T. Urata, K. Tanigaki, and T. Takahashi, Phys. Rev. Lett. {\bf113}, 237001 (2014).

\bibitem{ARPES3} P. Zhang, T. Qian, P. Richard, X. P. Wang, H. Miao, B. Q. Lv, B. B. Fu, T. Wolf, C. Meingast, X. X. Wu, Z. Q. Wang, J. P. Hu, and H. Ding, Phys. Rev. B {\bf91}, 214503 (2015).

\bibitem{ARPES4} M. D. Watson, T. K. Kim, A. A. Haghighirad, N. R. Davies, A. McCollam, A. Narayanan, S. F. Blake, Y. L. Chen, S. Ghannadzadeh, A. J. Schofield, M. Hoesch, C. Meingast, T. Wolf, and A. I. Coldea, Phys. Rev. B {\bf91}, 155106 (2015).

\bibitem{QuanOs1} T. Terashima, N. Kikugawa, A. Kiswandhi, E. S. Choi, J. S. Brooks, S. Kasahara, T. Watashige, H. Ikeda, T. Shibauchi, Y. Matsuda, T. Wolf, A. E. B\"{o}hmer, F. Hardy, C. Meingast, H. v. L\"{o}hneysen, M. T. Suzuki, R. Arita, and S. Uji, Phys. Rev. B {\bf90}, 144517 (2014).

\bibitem{QuanOs2} M. D. Watson, T. Yamashita, S. Kasahara, W. Knafo, M. Nardone, J. Beard, F. Hardy, A. McCollam, A. Narayanan, S. F. Blake, T. Wolf, A. A. Haghighirad, C. Meingast,A. J. Schofield, H. v. L\"{o}hneysen, Y. Matsuda, A. I. Coldea, and T. Shibauchi, Phys. Rev. Lett. {\bf115}, 027006 (2015).

\bibitem{ChengJGPressure} J. P. Sun, K. Matsuura, G. Z. Ye, Y. Mizukami, M. Shimozawa, K. Matsubayashi, M. Yamashita, T. Watashige, S. Kasahara, Y. Matsuda, J. Q. Yan, B. C. Sales, Y. Uwatoko, J. G. Cheng, and T. Shibauchi, Nat. Commun. {\bf7}, 12146 (2016).

\bibitem{CanfieldPressure} U. S. Kaluarachchi, V. Taufour, A. E. B\"{o}hmer, M. A. Tanatar, S. L. Bud¡¯ko, V. G. Kogan, R. Prozorov, and P. C. Canfield, Phys. Rev. B {\bf93}, 064503 (2016).

\bibitem{YuWQPRL} P. S. Wang, S. S. Sun, Y. Cui, W. H. Song, T. R. Li, R. Yu, H. C. Lei, and W. Q. Yu, Phys. Rev. Lett. {\bf117}, 237001 (2016).

\bibitem{HanaguriPRX} T. Watashige, Y. Tsutsumi, T. Hanaguri, Y. Kohsaka, S. Kasahara, A. Furusaki, M. Sigrist, C. Meingast, T. Wolf, H. v. L\"{o}hneysen, T. Shibauchi, and Y. Matsuda, Phys. Rev. X {\bf5}, 031022 (2015).

\bibitem{TailleferPRL} P. Bourgeois-Hope, S. Chi, D. A. Bonn, R. Liang, W. N. Hardy, T. Wolf, C. Meingast, N. Doiron-Leyraud, and L. Taillefer, Phys. Rev. Lett. {\bf117}, 097003 (2016).

\bibitem{FeSeCap} L. Jiao, C. L. Huang, S. R\"{o}{\ss}ler, C. Koz, U. K. R\"{o}{\ss}ler, U. Schwarz, and S. Wirth, Sci. Rep. {\bf7}, 44024 (2017).

\bibitem{ChenGY2017} G. Y. Chen, X. Y. Zhu, H. Yang, and H. H. Wen, arXiv:1703.08680v1 (2017).

\bibitem{DavisQPI} P. O. Sprau, A. Kostin, A. Kreisel, A. E. B\"{o}hmer, V. Taufour, P. C. Canfield, S. Mukherjee, P. J. Hirschfeld, B. M. Andersen, and J. C. S. Davis, Science {\bf357}, 75 (2017).

\bibitem{CuNernstReview} Y. Wang, L. Li, and N. P. Ong, Phys. Rev. B {\bf73}, 024510 (2006).

\bibitem{CuprateReview} A. Kaminski, T. Kondo, T. Takeuchi, and G. D. Gu, Philos. Mag. {\bf95}, 453 (2015).

\bibitem{YBCONernst} R. Daou, J. Chang, D. LeBoeuf, O. Cyr-Choini\`{e}re, F. Lalibert\'{e}, N. Doiron-Leyraud, B. J. Ramshaw, R. Liang, D. A. Bonn, W. N. Hardy, and L. Taillefer, Nature {\bf463}, 519 (2010).

\bibitem{YBCOMag} J. F. Yu, B. J. Ramshaw, I. Kokanovi\'{c}, K. A. Modic, N. Harrison, J. Day, R. Liang, W. N. Hardy, D. A. Bonn, A. McCollam, S. R. Julian, and J. R. Cooper, Phys. Rev. B {\bf92}, 180509(R) (2015).

\bibitem{SCFLaOFeAs} Z. W. Zhu, Z. A. Xu, X. Lin, G. H. Cao, C. M. Feng, G. F. Chen, Z. Li, J. L. Luo, and N. L. Wang, New J. Phys. {\bf10}, 063021 (2008).

\bibitem{SCFBaFeCoAs} G. Sheet, M. Mehta, D. A. Dikin, S. Lee, C. W. Bark, J. Jiang, J. D. Weiss, E. E. Hellstrom, M. S. Rzchowski, C. B. Eom, and V. Chandrasekhar, Phys. Rev. Lett. {\bf105}, 167003 (2010).

\bibitem{SCFFeSeTe} A. Pourret, L. Malone, A. B. Antunes, C. S. Yadav, P. L. Paulose, B. Fauqu\'{e}, and K. Behnia, Phys. Rev. B {\bf83}, 020504 (R) (2011).

\bibitem{SCFBaFeAsP} A. Ramos-\'{A}lvarez, J. Mosqueira, F. Vidal, D. Hu, G. F. Chen, H. Q. Luo, and S. L. Li, Phys. Rev. B {\bf92}, 094508 (2015).

\bibitem{SCFLiFeAs} F. Rullier-Albenque, D. Colson, A. Forget, and H. Alloul, Phys. Rev. Lett. {\bf109}, 187005 (2012).

\bibitem{ThermoReview} I. Pallecchi, F. Caglieris, and M. Putti, Supercond. Sci. Technol. {\bf29}, 073002 (2016).

\bibitem{SCFCsFeAs} H. Yang, J. Xing, Z. Y. Du, X. Yang, H. Lin, D. L. Fang, X. Y. Zhu, and H. H. Wen, Phys. Rev. B {\bf93}, 224516 (2016).

\bibitem{MatsudaPNAS} S. Kasaharaa, T. Watashigea, T. Hanagurib, Y. Kohsakab, T. Yamashitaa, Y. Shimoyamaa, Y. Mizukamia, R. Endoa, H. Ikedaa, K. Aoyamaa, T. Terashimae, S. Ujie, T. Wolff, H. von L\"{o}hneysenf, T. Shibauchia, and Y. Matsuda, Proc. Natl. Acad. Sci. USA {\bf111}, 16309 (2014).

\bibitem{MatsudaNC} S. Kasahara, T. Yamashita, A. Shi, R. Kobayashi, Y. Shimoyama, T. Watashige, K. Ishida, T. Terashima, T. Wolf, F. Hardy, C. Meingast, H. v. L\"{o}hneysen, A. Levchenko, T. Shibauchi, and Y. Matsuda, Nat. Commun. {\bf7}, 12843 (2016).

\bibitem{AEBohmer2013PRB} A. E. B\"{o}hmer, F. Hardy, F. Eilers, D. Ernst, P. Adelmann, P. Schweiss, T. Wolf, and C. Meingast, Phys. Rev. B {\bf87}, 180505 (R) (2013).

\bibitem{WHH} R. Werthamer, E. Helfand, and P. C. Hohenberg, Phys. Rev. {\bf147}, 295 (1966).

\bibitem{Tesanovic} Z. Te\v{s}anovi\'{c}, L. Xing, L. Bulaevskii, Q. Li, and M. Suenaga, Phys. Rev. Lett. {\bf69}, 3563 (1992)

\bibitem{GaoHPRB} H. Gao, C. Ren, L. Shan, Y. Wang, Y. Z. Zhang, S. P. Zhao, X. Yao, and H. H. Wen, Phys. Rev. B {\bf74}, 020505 (R) (2006).

\bibitem{WenHHPRL1999}  H. H. Wen, W. L. Yang, Z. X. Zhao, and Y. M. Ni, Phys. Rev. Lett. {\bf82}, 410 (1999).

\bibitem{J.Y.LinPRB2011} J. Y. Lin, Y. S. Hsieh, D. A. Chareev, A. N. Vasiliev, Y. Parsons, and H. D. Yang, Phys. Rev. B {\bf84}, 220507 (R) (2011).

\bibitem{Emery1995Nature} V. J. Emery, and S. A. Kivelson, Nature (London) {\bf374}, 434 (1995).

\bibitem{Wen2009PRL} H. H. Wen, G. Mu, H. Q. Luo, H. Yang, L. Shan, C. Ren, P. Cheng, J. Yan, and L. Fang, Phys. Rev. Lett. {\bf103}, 067002 (2009).

\bibitem{Alain1999PhysicaC} A. Junod, A. Erb, and C. Renner, Physica C {\bf317-318}, 333 (1999).

\bibitem{lambdaHe4} M.J. Buckingham, Prog. Low Temp. Phys. {\bf3}, 80 (1961).

\bibitem{Maki} K. Maki, Physica {\bf55}, 124 (1971).

\bibitem{BehniaReview} K. Behnia, and H. Aubin, Rep. Prog. Phys. {\bf79}, 046502 (2016).

\bibitem{BehniaNbSe2} R. Bel, K. Behnia, and H. Berger, Phys. Rev. Lett. {\bf91}, 066602 (2003).

\bibitem{NernstLaFeAsO} Q. Tao, Z. W. Zhu, X. Lin, G. H. Cao, Z. A. Xu, G. F. Chen, J. L. Luo, and N. L. Wang, J. Phys.: Condens. Matter {\bf22}, 072201 (2010).

\bibitem{NernstCaFe2As2} M. Matusiak, Z. Bukowski, and J. Karpinski, Phys. Rev. B {\bf81}, 020510 (R) (2010).

\bibitem{NernstEuFe2As2} M. Matusiak, Z. Bukowski, and J. Karpinski, Phys. Rev. B {\bf83}, 224505 (2011).

\bibitem{PengchengLi2007PRB} P. C. Li and R. L. Greene, Phys. Rev. B {\bf76}, 174512 (2007).

\bibitem{BlatterReview} G. Blatter, M. V. Feigel'man, V. B. Geshkenbein, A. I. Larkin, and V. M. Vinokur, Rev. Mod. Phys. {\bf66}, 1125 (1994).

\bibitem{LobbPRB1987} C. J. Lobb, Phys. Rev. B {\bf36}, 3930 (1987).

\bibitem{Qiangli1993PRB} Qiang Li, K. Shibutani, M. Suenaga, I. Shigaki, and R. Ogawa, Phys. Rev. B {\bf48}, 9877 (1993).

\bibitem{EFSC} G. P. Malik, J. Supercond. Nov. Magn. {\bf29}, 2755 (2016).

\bibitem{SCPara} I. Pallecchi, M. Tropeano, G. Lamura, M. Pani, M. Palombo, A. Palenzona, and M. Putti, Physica C {\bf482}, 68 (2012).

\bibitem{Orlando} T. P. Orlando, K. A. Delin, S. Foner, E. J. McNiff, Jr., J. M. Tarascon, L. H. Greene, W. R. McKinnon, and G. W. Hull, Phys. Rev. B {\bf36}, 2394 (R) (1987).

\bibitem{Nguyen} P. P. Nguyen, Z. H. Wang, A. M. Rao, M. S. Dresselhaus, J. S. Moodera, G. Dresselhaus, H. B. Radousky, R. S. Glass, and J. Z. Liu, Phys. Rev. B {\bf48}, 1148 (1993).

\bibitem{Zimmermann} P. Zimmermann, H. Keller, S. L. Lee, I. M. Savic, M. Warden, D. Zech, R. Cubitt, E. M. Forgan, E. Kaldis, J. Karpinski, and C. Kr\"{u}ger, Phys. Rev. B {\bf52}, 541 (1995).

\bibitem{MgB2Review} C. Buzea, and T. Yamashita, Supercond. Sci. Technol. {\bf14}, R115 (2001).

\bibitem{Caplin} A. D. Caplin, Y. Bugoslavsky, L. F. Cohen, L. Cowey, J. Driscoll, J. Moore, and G. K. Perkins, Supercond. Sci. Technol. {\bf16}, 176 (2002).

\bibitem{H.Yang} H. Yang, H. Luo, Z. Wang, and H. H. Wen, Appl. Phys. Lett., {\bf93}, 142506 (2008).

\bibitem{C.Ren} C. Ren, Z. S. Wang, H. Q. Luo, H. Yang, L. Shan, and H. H. Wen, Physica C {\bf469}, 599 (2009).

\bibitem{Tarantini} C. Tarantini, A. Gurevich, J. Jaroszynski, F. Balakirev, E. Bellingeri, I. Pallecchi, C. Ferdeghini, B. Shen, H. H. Wen, and D. C. Larbalestier, Phys. Rev. B {\bf84}, 184522 (2011).

\bibitem{ARPESBa122} H. Ding, K. Nakayama, P. Richard, S. Souma, T. Sato, T. Takahashi, M. Neupane, Y. M. Xu, Z. H. Pan, A. V. Federov, Z. Wang, X. Dai, Z. Fang, G. F. Chen, J. L. Luo, and N. L. Wang, J. Phys.: Condens. Matter, {\bf 23}, 135701 (2008).

\bibitem{Terashima} T. Terashima, N. Kikugawa, A. Kiswandhi, E. S. Choi, J. S. Brooks, S. Kasahara, T. Watashige, H. Ikeda, T. Shibauchi, Y. Matsuda, T. Wolf, A. E. B\"{o}hmer, F. Hardy, C. Meingast, H. v. L\"{o}hneysen, M. T. Suzuki, R. Arita, and S. Uji, Phys. Rev. B {\bf90}, 144517 (2014).

\bibitem{Hafiez} M. Abdel-Hafiez, J. Ge, A. N. Vasiliev, D. A. Chareev, J. Van de Vondel, V. V. Moshchalkov, and A. V. Silhanek, Phys. Rev. B {\bf88}, 174512 (2013).

\bibitem{FluctuationSC} A. Larkin and A. Varlamov, arXiv:cond-mat/0109177v1 (2001).

\bibitem{F.Pistolesi} F. Pistolesi and G. C. Strinati, Phys. Rev. B {\bf49}, 6356 (1994).

\bibitem{SrTiO3Marel} D. van der Marel, J. L. M. van Mechelen, and I. I. Mazin, Phys. Rev. B {\bf84}, 205111 (2011).

\bibitem{Uemura} Y. J. Uemura, Physica C {\bf 282-287}, 194 (1997).

\bibitem{BehniaSrTiO3} X. Lin, Z. W. Zhu, B. Fauqu\'{e}, and K. Behnia, Phys. Rev. X {\bf3}, 021002 (2013).

\bibitem{Ba122SH} J. G. Storey, J. W. Loram, J. R. Cooper, Z. Bukowski, and J. Karpinski, Phys. Rev. B {\bf88}, 144502 (2013).

\bibitem{FeImpurityPan} J. X. Yin, Z. Wu, J. H. Wang, Z. Y. Ye, J. Gong, X. Y. Hou, L. Shan, A. Li, X. J. Liang, X. X. Wu, J. Li, C. S. Ting, Z. Q. Wang, J. P. Hu, P. H. Hor, H. Ding, and S. H. Pan, Nat. Phys. {\bf11}, 543 (2015).


\end{thebibliography}
\end{document}